\documentstyle[sprocl,epsf,feynmp,epsfig]{article}

\def\cal{\mathcal}
\def\pls{\makebox(0,0){$+$}}
\def\mins{\makebox(0,0){$-$}}
\def\ssp{\makebox(0,0)
    {\thinlines\put(-.1,0){\line(1,0){.2}}\put(0,-.1){\line(0,0){.2}}}}
\def\ssm{\makebox(0,0){\put(-.1,0){\thinlines\line(1,0){.2}}}}
\def\loopa{
\unitlength 1.00mm
\thicklines
\begin{picture}(10.00,8.93)
\bezier{28}(2.00,5.00)(2.13,8.33)(6.00,8.93)
\bezier{28}(2.00,4.87)(2.13,1.53)(6.00,0.93)
\bezier{28}(10.00,5.00)(9.87,8.33)(6.00,8.93)
\bezier{28}(10.00,4.87)(9.87,1.53)(6.00,0.93)
\bezier{28}(2.00,-3.00)(2.13,0.33)(6.00,0.93)
\bezier{28}(2.00,-3.13)(2.13,-6.47)(6.00,-7.07)
\bezier{28}(10.00,-3.00)(9.87,0.33)(6.00,0.93)
\bezier{28}(10.00,-3.13)(9.87,-6.47)(6.00,-7.07)
\put(6.00,1.00){\circle*{1.00}}
\put(6.40,8.80){\vector(1,0){0.2}}
\put(5.53,8.73){\line(1,0){0.87}}
\put(5.53,-6.93){\vector(-1,0){0.2}}
\put(6.47,-7.00){\line(-1,0){0.93}}
\end{picture}}
\def\loopb#1{
\unitlength 1.00mm
\thicklines
\begin{picture}(17.50,13.00)
\put(2.00,1.00){\circle*{1.00}}
\put(17.00,1.00){\circle*{1.00}}
\bezier{68}(2.00,1.00)(9.50,5.00)(17.00,1.00)
\bezier{68}(2.00,1.00)(9.50,-3.00)(17.00,1.00)
\bezier{112}(2.00,1.00)(9.50,13.00)(17.00,1.00)
\bezier{112}(2.00,1.00)(9.50,-11.00)(17.00,1.00)
\put(9.93,-1.07){\vector(1,0){0.2}}
\put(9.00,-1.07){\line(1,0){0.93}}
\put(9.93,2.93){\vector(1,0){0.2}}
\put(8.93,2.93){\line(1,0){1.00}}
\put(9.13,6.93){\vector(-1,0){0.2}}
\put(10.07,6.93){\line(-1,0){0.93}}
\put(9.20,-5.00){\vector(-1,0){0.2}}
\put(10.07,-5.07){\line(-1,0){0.87}}
\if#1c \multiput(9.5,-8.)(0,6.5){3}{\line(0,1){5}}\fi
\end{picture}}
\def\loopc#1{ 
\unitlength 1.00mm
\thicklines
\begin{picture}(18.54,9.47)
\put(2.00,-5.00){\circle*{0.93}}
\put(18.00,-5.00){\circle*{1.07}}
\put(10.00,9.00){\circle*{0.94}}
\bezier{72}(2.00,-5.00)(10.00,-1.00)(18.00,-5.00)
\bezier{72}(2.00,-5.00)(10.00,-9.00)(18.00,-5.00)
\bezier{80}(2.00,-5.00)(2.00,6.00)(10.00,9.00)
\bezier{72}(2.00,-5.00)(9.07,-0.80)(10.00,9.00)
\bezier{72}(18.00,-5.00)(11.07,-0.93)(10.00,9.00)
\bezier{76}(18.00,-5.00)(18.13,5.07)(10.00,9.00)
\put(9.93,-7.00){\vector(-1,0){0.2}}
\put(11.00,-7.00){\line(-1,0){1.07}}
\put(3.80,3.27){\vector(1,2){0.2}}
\multiput(3.40,2.47)(0.10,0.20){4}{\line(0,1){0.20}}
\put(16.73,2.33){\vector(2,-3){0.2}}
\multiput(16.20,3.13)(0.11,-0.16){5}{\line(0,-1){0.16}}
\put(10.07,-3.00){\vector(1,0){0.2}}
\put(9.07,-3.00){\line(1,0){1.00}}
\put(12.53,0.60){\vector(-2,3){0.2}}
\multiput(13.00,-0.20)(-0.12,0.20){4}{\line(0,1){0.20}}
\put(7.60,0.67){\vector(-2,-3){0.2}}
\multiput(8.00,1.33)(-0.10,-0.17){4}{\line(0,-1){0.17}}
\if#1c \multiput(8.5,-9.5)(3.5,5.25){3}{\thinlines\line(2,3){2.5}}\fi
\end{picture}}
\def\ssp{\makebox(0,0)
  {\thicklines\put(-.1,0){\line(1,0){.2}}\put(0,-.1){\line(0,0){.2}}}}
\def\ssm{\makebox(0,0){\put(-.1,0){\thicklines\line(1,0){.2}}}}
\def\phidecomposition{
     \put(2,0){\thicklines\oval(3.0,1.8)[l]
     \put(-.7,0){\makebox(0,0){$\alpha$}}}
     \put(3,0){\thicklines\oval(3.0,1.8)[r]
     \put(.8,0){\makebox(0,0){$\beta$}}}
     \thicklines
     \put(2,-.9){\line(0,1){1.8}}
     \put(3,-.9){\line(0,1){1.8}}
     \put(2.5,0){\thicklines
     \multiput(0,0.25)(0,0.22){3}{\put(-.5,0){\vector(1,0){1}}
                            \put(-.7,0){\ssp}
                            \put(0.7,0){\ssm} }
     \multiput(0,-.25)(0,-.22){3}{\put(0.5,0){\vector(-1,0){1}}
                            \put(-.7,0){\ssp}
                            \put(0.7,0){\ssm} }}}
\def\wa#1{
\unitlength 0.7mm
\thicklines
\begin{picture}(7,5)\put(2,0){
\put(2.00,1.00){\circle*{1.00}}
\put(2.00,6.00){\makebox(0,0){$ #1 $}}
\bezier{16}(0.54,-0.41)(3.54,2.55)(3.54,2.55)
\bezier{16}(0.54,2.55)(3.50,-0.41)(3.50,-0.41)
\put(1.21,0.22){\vector(1,1){0.2}}
\multiput(0.54,-0.45)(0.11,0.11){6}{\line(0,1){0.11}}
\put(4.61,3.51){\vector(1,1){0.8}}
\multiput(2.32,1.22)(0.11,0.11){20}{\line(0,1){0.11}}
\put(1.25,1.76){\vector(3,-4){0.2}}
\multiput(0.58,2.55)(0.11,-0.13){6}{\line(0,-1){0.13}}
\put(4.54,-1.37){\vector(1,-1){0.8}}
\multiput(2.25,0.76)(0.13,-0.12){18}{\line(1,0){0.13}}
}
\end{picture}}
\def\wb#1{
\unitlength 0.7mm
\thicklines
\begin{picture}(17.51,9.01)\put(4,0){
\put(9.51,-6.49){\circle*{1.00}}
\put(9.51,8.51){\circle*{1.00}}
\put(9.51,-9.49){\makebox(0,0){$ #1 $}}
\put(9.51,11.51){\makebox(0,0){$ #1 $}}
\bezier{68}(9.51,-6.49)(5.51,1.01)(9.51,8.51)
\bezier{68}(9.51,-6.49)(13.51,1.01)(9.51,8.51)
\put(11.58,1.44){\vector(0,1){0.2}}
\put(11.58,0.51){\line(0,1){0.93}}
\put(7.52,0.58){\vector(0,-1){0.2}}
\put(7.52,1.51){\line(0,-1){0.93}}
\bezier{20}(7.00,8.55)(12.04,8.55)(12.04,8.55)
\bezier{20}(7.00,-6.49)(12.04,-6.53)(12.04,-6.53)
\put(8.00,8.51){\vector(1,0){0.2}}
\put(7.00,8.51){\line(1,0){1.00}}
\put(13.44,8.55){\vector(1,0){0.2}}
\put(12.44,8.55){\line(1,0){1.00}}
\put(8.00,-6.58){\vector(1,0){0.2}}
\put(7.00,-6.58){\line(1,0){1.00}}
\put(13.44,-6.58){\vector(1,0){0.2}}
\put(12.44,-6.58){\line(1,0){1.00}}
}
\end{picture}
}
\newcommand{\di}{{\mathrm d}}
\newcommand{\Tr}{{\mathrm{Tr}}}
\newcommand{\ii}{{\mathrm i}}

\renewcommand{\and}{\quad{\mathrm{and}}\quad}

\renewcommand{\Re}{{\mathrm{Re}}}

\renewcommand{\Im}{{\mathrm{Im}}}

\def\scr#1{\mbox{\scriptsize #1}}

\def\vec#1{\mbox{\boldmath $#1$}}
\newcommand{\dpi}[1]{\frac{\di^4 #1}{(2\pi)^4}}                
\newcommand{\Pbr}[1]{\left\{#1\right\}}                    
\newlength{\charwidth}

\newcommand{\lap}%
{\raisebox{-0.5ex}{$\stackrel{\scriptstyle <}{\scriptstyle \sim}$}}
\newcommand{\gap}%
{\raisebox{-0.5ex}{$\stackrel{\scriptstyle >}{\scriptstyle \sim}$}}

\def\Gr{G}\def\Se{\Sigma}

\def\Ga{\Gr}
\def\Sa{\Se}

\def\Lg{{\cal L}}
\def\Lgh{\makebox[3.5mm]{${\widehat{\makebox[2mm]{$\Lg$}}}$}\vphantom{L}}
\def\Lint{\Lgh^{\mbox{\scriptsize int}}}

\def\A{A}
\def\Gm{\Gamma}
\def\F{F}                             
\def\Ft{\widetilde{F}}                 
\def\Fd{F}                             
\def\Fdt{\widetilde{F}}                
\def\fd{f}                             

\def\tp{\widetilde{p}}\def\tm{\widetilde{m}}\def\tW{\widetilde{W}} 
\def\Get{\Gamma_{\scr{out}}}   
\def\Gbt{\Get}                                   
\def\Ge{\Gamma_{\scr{in}}}     
\def\Gb{\Ge}              

\def\Ld{\Gamma_{\scr{out}}}
\def\Ldt{\Gamma_{\scr{in}}}
\def\vu{v}

\def\Do{{\cal D}}

\def\oneloop{
     \put(1.5,0){\thicklines\oval(2.0,1.5)}
     \put(0,0){\photon}\put(0.3,0.3){\ssp} 
     \put(2.5,0){\photon}\put(2.7,0.3){\ssm}}
\def\hatchedself{\begin{picture}(5,1)\put(0,0){\oneloop}
     \multiput(0.95,0)(.2,0){5}{\put(0,-.7){\line(1,4){.35}}
     \put(0,.7){\line(1,-4){.35} }}
     \end{picture}}
\def\interaction{\makebox(0,0){\put(0,0){\interact}
    \put(0,.95){\ssp}\put(0,-.95){\ssm}
    \put(0,.125){\ssp}\put(0,-.125){\ssm}}}
\def\interact{\makebox(0,0){\put(0,.5){\fullbox}
    \thicklines\put(0,0){\oval(.75,.5)}
    \put(0,-.5){\fullbox}} }
\def\til2loop{\put(0,0){\oneloopvertex}\put(4.,0){\pls}
      \put(4.75,0){\oneloopvertex}\put(6.375,0){\interaction}
      \put(9,0){\pls}}
\def\classdiagram{
     \begin{picture}(18,3)
     \put(0,0){\til2loop}\put(11,0){\nloop}
     \put(10,0){\makebox(0,0){$\dots$}}
     \put(18,0){\makebox(0,0){$\dots$}}  
     \end{picture}}
\def\photon{\thinlines\multiput(0,0)(.2,0){3}{\line(1,0){0.1}}}
\def\oneloopvertex{
    \put(1.625,0){\thicklines\oval(2.0,1.5)}
    \put(0,0){\photon}\put(0.35,0.3){\ssp} 
    \put(0.625,0){\circle*{.25}}\put(2.625,0){\circle*{.25}}
    \put(2.75,0){\photon}\put(2.9,0.3){\ssm}}
\def\fullbox{\makebox(0,0){\rule{1.5mm}{3mm}}}
\def\nloop{\begin{picture}(6.5,.5) \put(0,0){\longloop}
       \multiput(1.625,0)(1,0){2}{\interaction}\put(3.625,0)
       {\makebox(0,0){$\dots$}}\put(4.625,0){\interaction}\end{picture}}
\def\longloop{
             \put(3.125,0){\thicklines\oval(5.0,1.5)}
             \put(0,0){\photon}\put(0.35,0.3){\ssp} 
             \put(5.75,0){\photon}\put(5.9,0.3){\ssm}
             \put(0.625,0){\circle*{.25}}\put(5.625,0){\circle*{.25}}}
\def\Boson{\thicklines
           \multiput(0.0625,0)(.25,0){6}{\oval(.125,.125)[b]}
           \multiput(0.1875,0)(.25,0){6}{\oval(.125,.125)[t]}
           }
\def\Grho{\Green\thicklines
        \put(0.75,0){\oval(1.5,0.3)}
        \put(0,0){\Black\circle*{0.2}}\put(1.5,0){\Black\circle*{0.2}}
        \put(0,0){\line(1,0){0.8}}\put(1.5,0){\vector(-1,0){1}}}
\def\Blue{}\def\blue{}
\def\Red{}\def\Green{}\def\Black{}
\def\citerange[#1-#2]{[\citem[#1,@]-\citem[#2,@]]}
\def\citem[#1,#2]{\csname b@#1\endcsname\if @#2{}\else ,\citem[#2]\fi}
\begin{document}

\title{SOFT MODES, QUANTUM TRANSPORT  AND  KINETIC ENTROPY$^{\S}$\footnotetext{$^{\S}$Combined contribution summarizing the talks of J. Knoll on
  the subject ``Soft Modes and Resonances in Dense Matter'' and Y.B. Ivanov on
  the subject ``Quantum Transport and Kinetic Entropy''.}}

\author{
YU. B. IVANOV$^\dag$\footnotetext{$^\dag$Permanent address: 
Kurchatov Institute, Kurchatov sq. 1, Moscow 123182, Russia}, 
J. KNOLL, H. VAN HEES, 
D. N. VOSKRESENSKY$^\ddag$\footnotetext{$^\ddag$Permanent address: 
Moscow Institute for Physics and Engineering, 
Kashirskoe sh. 31, Moscow 115409, Russia}}



\address{Gesellschaft f\"ur Schwerionenforschung mbH, Planckstr. 1,
64291 Darmstadt, Germany\\
E-mail: Y.Ivanov@gsi.de, J.Knoll@gsi.de, D.Voskresensky@gsi.de}



\maketitle\abstracts{ The effects of the propagation of particles which have a
  finite life-time and an according width in their mass spectrum are discussed
  in the context of transport descriptions. In the first part the coupling of
  soft photon modes to a source of charged particles is studied in a classical
  model which can be solved completely in analytical terms. The solution
  corresponds to a re-summation of certain field theory diagrams. The general
  properties of broad resonances in dense finite temperature systems are
  discussed at the example of the $\rho$-meson in hadronic matter. The second
  part addresses the problem of transport descriptions which also account for
  the damping width of the particles. The Kadanoff--Baym equation after
  gradient approximation together with the $\Phi$-derivable method of Baym
  provides a self-consistent and conserving scheme. Memory effects appearing in
  collision term diagrams of higher order are discussed.  We derive a
  generalized expression for the nonequilibrium kinetic entropy flow, which
  includes corrections from fluctuations and mass-width effects. In special
  cases an $H$-theorem is proved. Memory effects in collision terms provide
  contributions to the kinetic entropy flow that in the Fermi-liquid case
  recover the famous bosonic type $T^3 \ln T$ correction 
  to the specific heat of liquid Helium-3. 
  }

With the aim to describe the collision of two nuclei at intermediate or even
high energies one is confronted with the fact that the dynamics has to include
particles like the delta or rho-meson resonances with life-times of less than
2 fm/c or equivalently with damping widths above 100 MeV. Also the collisional
damping rates deduced from presently used transport codes are of the same 
order, whereas typical mean temperatures 
range between 50 to 150 MeV depending on beam energy. Thus, the damping width
of most of the constituents in the system can by no means be treated as a
perturbation.  As a consequence the mass spectrum of the particles in the
dense matter is no longer a sharp delta function but rather acquires a width
due to collisions and decays. One thus comes to a picture which unifies {\em
  resonances} which have already a width in vacuum due to decay modes with the
''states'' of particles in dense matter, which obtain a width due to
collisions (collisional broadening). 

The theoretical concepts for a proper many body description in terms of a real
time nonequilibrium field theory have already been devised by
Schwinger~\cite{Schw}, Kadanoff and Baym~\cite{Kad62}, and
Keldysh~\cite{Keld64} in the early sixties, 
extensions to relativistic plasmas followed by 
Bezzerides and DuBois~\cite{Bez}. 
First investigations of the quantum effects on the Boltzmann
collision term were given by Danielewicz~\cite{D}, the principal conceptual
problems on the level of quantum field theory were investigated by
Landsmann~\cite{Landsmann}, while applications which seriously include the
finite width of the particles in transport descriptions were carried out only
in recent times, e.g. refs.~\cite{D,DB,BM,HFN,PH,QH,Weinhold,KV}.  For
resonances, 
e.g. the delta resonance, it was natural to consider broad mass distributions
and ad hoc recipes have been invented to include this in transport simulation
models.  However, many of these recipes are not correct as they violate some
basic principles like detailed balance (see discussion in
ref.~\cite{DB}), and the description of
resonances in dense matter has to be improved~\cite {Weinhold}.

In this contribution the consequences of the propagation of particles with
short life-times are discussed. First we address a genuine
problem related to the occurrence of broad damping width, i.e. the soft mode
problem. At the classical level we investigate the coupling of a
coherent classical field, the Maxwell field, to a stochastic source
described by the Brownian motion of a charged particle. In this case the
classical current-current correlation function, can be obtained in closed
analytical terms and discussed as a function of the macroscopic transport
properties, the friction and diffusion coefficient of the Brownian particle.
The result corresponds to a partial re-summation of photon self-energy
diagrams in the real-time formulation of field theory. Subsequently the
properties of particles with broad damping width is illustrated at the example
of the $\rho$-meson in dense matter at finite temperature.  In the final part
we discuss how particles with such broad mass-width can be described
consistently within a transport theoretical picture.

We are going to argue that the Kadanoff--Baym equations in the first gradient
approximation together with the $\Phi$-functional method of Baym~\cite{Baym}
provide a proper frame for kinetic description of systems of particles with a
broad mass-width. To this end, we discuss relevant problems concerning charge
and energy--momentum conservation, thermodynamic consistency, memory effects
in the collision term and the growth of entropy in specific
cases.  For simplicity we concentrate on systems of non-relativistic
particles.  Generalization to systems of relativistic particles and bosonic
mean fields can be straight forwardly done along the lines given in
ref.~\cite{IKV}. 

\section{Bremsstrahlung from Classical Sources}
For a clarification of the soft mode problem we discus an example in classical
electrodynamics. We consider a stochastic source, the hard matter, where the
motion of a single charge is described by a diffusion process in terms of a
Fokker-Planck equation for the probability distribution $f$ of
position ${\vec x}$
and velocity ${\vec v}$
\begin{eqnarray}\label{FP}
    \frac{\partial}{\partial t} f({\vec x},{\vec v},t)
    =
    \left({D\Gamma_x^2}\frac{\partial^2}{\partial {\vec v}^2}
    +\Gamma_x \frac{\partial}{\partial {\vec v}}{\vec v}-{\vec
    v}\frac{\partial}{\partial {\vec x}}\right) f({\vec x},{\vec v},t).
\end{eqnarray}
Fluctuations also evolve in time according to this equation, or
equivalently by a random walk process~\cite{KV}, and this way determine
correlations. This
charge is coupled to the Maxwell field. On the assumption of a non-relativistic
source, this case does not suffer from standard pathologies encountered in hard
thermal loop (HTL) problems of QCD, namely the collinear singularities, where
${\vec v}{\vec q}\approx 1$, and from diverging Bose-factors. The advantage of
this Abelian example is that damping can be fully included without violating
current conservation and gauge invariance. This problem
is related to the Landau--Pommeranchuk--Migdal effect of bremsstrahlung in
high-energy scattering~\cite{LPM}.
\begin{figure}
\parbox{5.6cm}{\epsfxsize=5.6cm\epsfbox{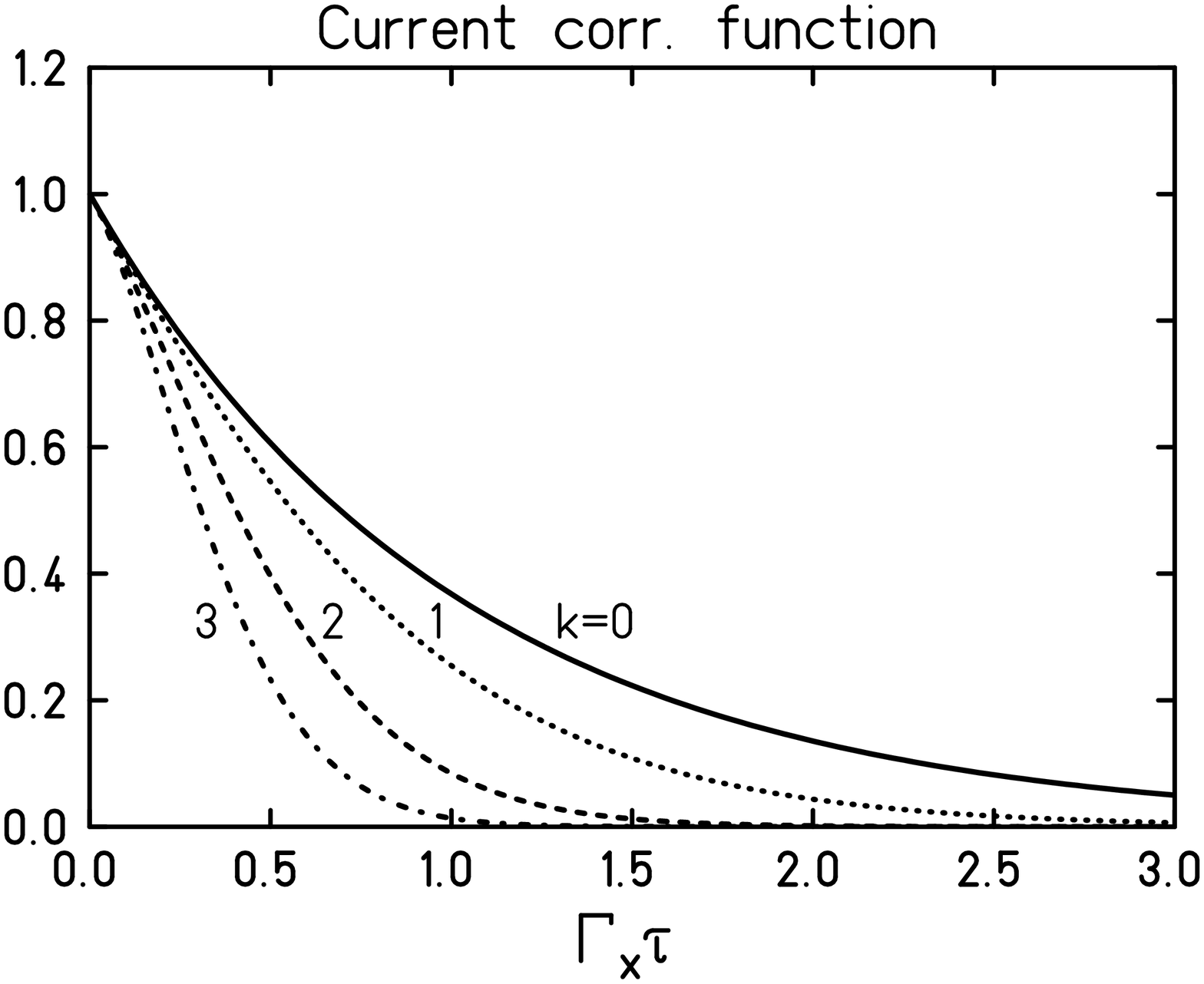}}\hfill
\parbox{5.6cm}{\epsfxsize=5.6cm\epsfbox{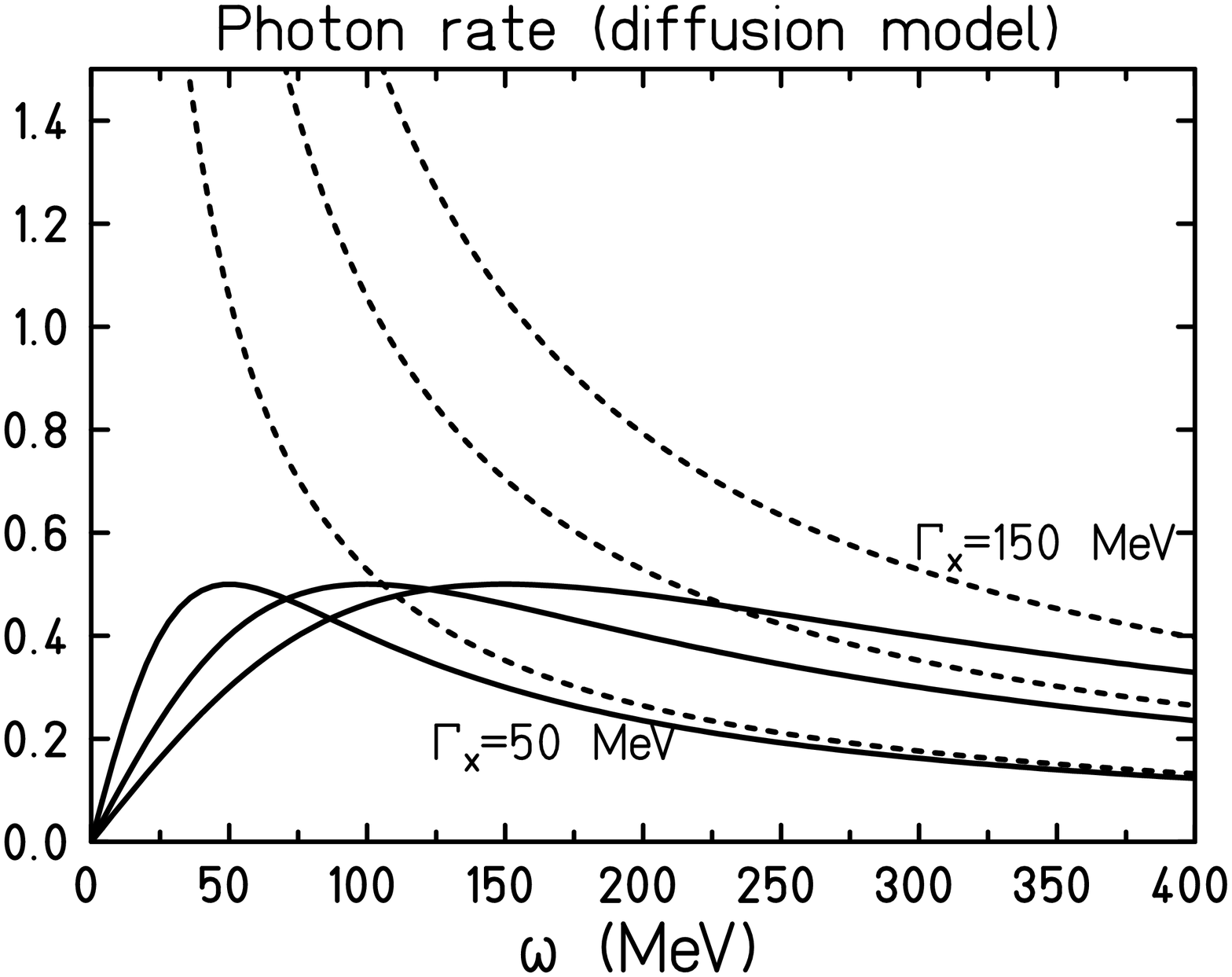}}
\caption{Left: Current-current correlation function
in units of $e^2<v^2>$ as a
function of time (in units of $1/\Gamma_x$) for different values of
the photon momentum $q^2=3k^2\Gamma_x^2/{<v^2>}$ with $k=0,1,2,3$.
Right: Rate of real photons $\di^2
N/(\di\omega\di t)$ in units of $4\pi e^2\left<{\vec v}^2\right>/3$
for a non-relativistic source for $\Gamma_x=$50,100,150 MeV; for
comparison the IQF results (dashed
lines) are also shown.}
\label{CCC}\label{Rate}
\end{figure}

The two macroscopic parameters, the spatial diffusion $D$ and friction 
$\Gamma_x$ coefficients 
determine the relaxation rates of velocities.  In the
equilibrium limit ($t\rightarrow\infty$) the distribution attains a
Maxwell-Boltzmann velocity distribution with the temperature $T=m\left< {\vec
    v}^2\right>/3=mD\Gamma_x$.  The correlation function can be obtained in
closed form and one can discuss the resulting time correlations of the current
at various values of the spatial photon momentum ${\vec q}$, Fig.
\ref{CCC} (details are given in ref.~\cite{KV}).  For the transverse part of
the correlation tensor this correlation decays exponentially as $\sim
e^{-\Gamma_x\tau}$ at ${\vec q}=0$, and its width further decreases with
increasing momentum $q=|{\vec q}|$. The in-medium production rate is given by
the time Fourier transform $\tau\rightarrow\omega$, Fig. \ref{Rate} (right
part). The hard part of the spectrum behaves as intuitively expected, namely,
it is proportional to the microscopic collision rate expressed through 
$\Gamma_x$ (cf. below) and thus can be treated pertubatively by incoherent
quasi-free (IQF) 
scattering prescriptions. However, independently of $\Gamma_x$ the rate
saturates at a value of $\sim 1/2$ in these units around $\omega\sim\Gamma_x$,
and the soft part shows the inverse behavior.  That is, with increasing
collision rate the production rate is more and more suppressed! This is in
line with the picture, where photons cannot resolve the individual
collisions any more. Since the soft part of the spectrum behaves like $
\omega/\Gamma_x$, it shows a genuine non-perturbative feature which cannot be
obtained by any power series in $\Gamma_x$.  For comparison: the dashed lines
show the corresponding IQF yields, which agree with the correct rates for the
hard part while completely fail and diverge towards the soft end of the
spectrum.  For non-relativistic sources $\left<{\vec v}^2\right>\ll 1$ one can
ignore the additional $q$-dependence (dipole approximation; cf. Fig.
\ref{CCC}) and the entire spectrum is determined by one macroscopic scale, the
relaxation rate $\Gamma_x$. This scale provides a quenching factor
\begin{equation}\label{suppression}
C_0(\omega)=\frac{\omega^2}{\omega^2+\Gamma_x^2}\; 
\end{equation}
by which the IQF results have to be corrected in order to account for
the finite collision time effects in dense matter.\\[2mm]

The diffusion result represents a re-summation of the microscopic Langevin
multiple collision picture and altogether only macroscopic scales are relevant
for the form of the spectrum and not the details of the microscopic
collisions. Note also that the classical result fulfill the
classical version ($\hbar\rightarrow 0$) of the sum rules discussed in 
refs.~\cite{DGK,KV}.

\section{Radiation on the Quantum level}
We have seen that at the classical level the problem of radiation from
dense matter can be solved quite naturally and completely at least for
simple examples, and Figs. \ref{CCC} 
display the main
physics. They show, that the {\em damping} of the particles due to
scattering is an important feature, which in particular has to be
included right from the onset.  This does not only assure results
that no longer diverge, but also provides a systematic and convergent
scheme. On the {\em quantum level} such problems require techniques beyond the
standard repertoire of perturbation theory or the quasi-particle
approximation.

\noindent\parbox[t]{9.5cm}{
In terms of nonequilibrium diagrammatic technique in Keldysh notation,
the production or 
  absorption rates are given by photon self-energy diagrams of the type to the
  right with an in-- and out-going photon line (dashed). The\hfill 
  hatched\hfill loop\hfill area\hfill
  }  \parbox[t]{2.6cm}{\unitlength7mm
  \begin{picture}(5,.7) \put(0.4,-.6){\hatchedself}
\end{picture}}\\[.5ex]
denotes all strong interactions of the source. The latter give rise to a
whole series of diagrams.  As mentioned, for the particles of the source, e.g.
the nucleons, one has to re-sum Dyson's equation with the corresponding full
complex self-energy in order to determine the full Green's functions in dense
matter. Once one has these Green's functions together with the interaction
vertices at hand, one could in principle calculate the required diagrams.
However, both the computational effort to calculate a single diagram and the
number of diagrams increase dramatically with the loop order of the
diagrams, such that in practice only lowest-order loop diagrams can be
considered in the quantum case. In certain limits some diagrams drop out.
We could show that in the {\em classical limit},
which in this case implies the hierarchy $\omega,|{\vec q}|,\Gamma\ll T\ll m$
together with low phase-space occupations for the source, i.e.  $f(x,p)\ll 1$,
only the following set of diagrams survives
\begin{equation}\label{classicaldiagram}
\unitlength6mm
\begin{picture}(18.5,1)\put(0,.2){\classdiagram}\end{picture}
\vphantom{\int_{\int}^{\int}}
\end{equation}
In these diagrams the bold lines denote the full nucleon Green's functions
which also include the damping width, the black blocks represent the effective
nucleon-nucleon interaction in matter, and the full dots the coupling vertex
to the photon. Each of these diagrams with $n$ interaction loop insertions
just corresponds to the $n^{th}$ term in the corresponding classical Langevin
process, where hard scatterings occur at random with a constant {\em mean
  collision rate} $\Gamma$. These scatterings consecutively change the
velocity of a point charge from ${\vec v}_0$ to ${\vec v}_1$ to ${\vec
  v}_2$, $\dots$. In between scatterings the charge moves freely. For such
a multiple collision process the space integrated current-current correlation
function takes a simple Poisson form
\begin{eqnarray}\label{Apoisson}
\ii\Pi^{\mu\nu -+}   &\propto&\int \di^3x_1\di^3x_2
\makebox[3.5cm][l]{$ \left<j^{\nu}({\vec
      x_1},t-\mbox{$\frac{\tau}{2}$})j^{\mu}({\vec 
 x_2},t+\mbox{$\frac{\tau}{2}$})\right>$}\cr
&=& e^2 
   \left<v^{\mu}(0)v^{\nu}(\tau)\right>
=
   e^2e^{-|\Gamma\tau|}\sum_{n=0}^\infty
   \frac{|\Gamma\tau|^n}{n!} \left< v^{\mu}_0 v^{\nu}_n\right>
\end{eqnarray} 
with $v=(1,{\vec v})$. Here $\left<\dots\right>$ denotes the average over
the discrete collision sequence. This form, which one writes down
intuitively, agrees with the analytic result of the quantum correlation
diagrams (\ref{classicaldiagram}) in the limit $n\ll 1$ and $\Gamma\ll
T$. 
Fourier
transformed it determines the spectrum in completely regular terms (void of
any infra-red singularities), where each term describes the interference of the
photon being emitted at a certain time or $n$ collisions later.  In special
cases where velocity fluctuations are degraded by a constant fraction $\alpha$
in each collision, such that $\left< {\vec v}_0\cdot{\vec v}_n \right>=
\alpha^n\left< {\vec v}_0\cdot{\vec v}_0 \right>$, one can re-sum the
whole series in Eq. (\ref{Apoisson}) and thus recover the relaxation result with
$2\Gamma_x\left< {\vec v}^2 \right>= \Gamma\left< ({\vec v}_0-{\vec
    v}_1)^2 \right>$ at least for ${\vec q}=0$ and the corresponding
quenching factor (\ref{suppression}). Thus the classical multiple collision
example provides a quite intuitive picture about such diagrams. Further
details are given in ref.~\cite{KV}.

The above example shows that we have to deal with particle transport
that explicitly takes account of the particle mass-width in order to
properly describe soft radiation from the system. 

\section{The $\rho$-meson in dense matter}

Another example we like to discuss concerns properties of the
$\rho$-meson and their consequences for the $\rho$-decay into
di-leptons are discussed.  In terms of the nonequilibrium diagrammatic
technique, the exact production rate of di-leptons 
is given by the following formula \unitlength6mm
\begin{eqnarray}\label{dndtdm}
\frac{\di n^{\mbox{e}^+\mbox{e}^-}}{\di t\di m}&=&
\begin{picture}(7.5,1.5)\put(0.,0.2){
        \put(1.5,0){\vector(-1,1){0.8}}\put(0.2,1.3){\line(1,-1){0.55}}
        \put(1.5,0){\vector(-1,-1){0.8}}\put(0.2,-1.3){\line(1,1){0.55}}
        \put(.8,1.2){\makebox(0,0){\small e$^+$}}
        \put(.8,-1.2){\makebox(0,0){\small e$^-$}}
        \put(1.5,0){\Boson}\put(2.25,-.7){\makebox(0,0){$\gamma^*$}}
        \put(3,0){\Grho}\put(3.75,-.7){\makebox(0,0){\Green$\rho$}}
        \put(3,0.5){\mins}\put(4.5,0.5){\pls}
        \put(4.5,0){\Boson}\put(5.25,-.7){\makebox(0,0){$\gamma^*$}}
        \put(6,0){\line(1,1){0.8}}\put(7.3,1.3){\vector(-1,-1){0.6}}
        \put(6,0){\line(1,-1){0.8}}\put(7.3,-1.3){\vector(-1,1){0.6}}
        \put(6.8,1.2){\makebox(0,0){\small e$^+$}}
        \put(6.8,-1.2){\makebox(0,0){\small e$^-$}}
        \put(1.5,0){\circle*{0.2}}
        \put(6,0){\circle*{0.2}}
        }
\end{picture}\nonumber\\[6mm]
&=&{\Green f_{\rho}(m,{\vec p},{\vec x},t)\;
        A_{\rho}(m,{\vec p},{\vec x},t)}\;
2m\;\Gamma^{\rho\;\mbox{\small e}^+\mbox{\small e}^-}(m).
\end{eqnarray}
Here $\Gamma^{\rho\;\mbox{\small e}^+\mbox{\small e}^-}(m)\propto1/m^3$ is the
mass-dependent electromagnetic decay rate of the $\rho$
into the 
di-lepton pair of invariant mass $m$. 
The phase-space distribution $f_{\rho}(m,{\vec p},{\vec
  x},t)$ and the spectral function $A_{\rho}(m,{\vec p},{\vec x},t)$ define
the properties of the $\rho$-meson at space-time point ${\vec x},t$. Both
quantities are in principle to be determined dynamically by an appropriate
transport model. However till to-date the spectral functions are not treated
dynamically in most of the present transport models. Rather one employs
on-shell $\delta$-functions for all stable particles and spectral functions
fixed to the vacuum shape for resonances.

As an illustration, the model case is discussed, where the $\rho$-meson just
strongly couples to two channels, i.e. the $\pi^+\pi^-$ and $\pi
N\leftrightarrow\rho N$ channels, the latter being relevant at finite nuclear
densities. The latter component is representative for all channels
contributing to the so-called {\em direct $\rho$} in transport codes. For a
first orientation the equilibrium properties\footnote{Far more sophisticated
  and in parts unitary consistent equilibrium calculations
have already been presented in the
literature~\cite{Rapp,Mosel,Klingl,FrimanPirner,FLW}.
It is not the point to compete with them at this
place.} are discussed in simple
analytical terms with the aim
to discuss the consequences for the implementation of such resonance processes
into dynamical transport simulation codes.

Both considered processes add to the total width of the $\rho$-meson
\begin{eqnarray}\label{Gammatot}
\Gamma_{\rm tot}(m,{\vec p})&=&\Gamma_{\rho\rightarrow{\pi}^+{\pi}^-}(m,{\vec
  p})+ 
\Gamma_{\rho\rightarrow{\pi} NN^{-1}}(m,{\vec p}),
\end{eqnarray}
and the equilibrium spectral function then results from the cuts of the two
diagrams \unitlength6mm
\begin{eqnarray}\label{A2}
{\Green A_{\rho}(m,{\vec p})}\;&=&\normalsize
\underbrace{
\begin{picture}(5.5,1.3)\thicklines\put(0.25,0.2){
        \put(0,0){\Grho}
        \put(3.5,0){\Grho}
        \put(2.5,0){\Blue\oval(2,1.5)}
        \put(2.6,0.75){\Blue\vector(-1,0){0.3}}
        \put(2.6,-0.75){\Blue\vector(-1,0){0.3}}
        \put(2.5,1.25){\makebox(0,0){\Blue$\pi^+$}}
        \put(2.5,-1.25){\makebox(0,0){\Blue$\pi^-$}}
        \put(1.95,-1.1){\thinlines\line(1,2){0.3}}
        \put(2.35,-0.3){\thinlines\line(1,2){0.3}}
        \put(2.75,.5){\thinlines\line(1,2){0.3}}
        }
\end{picture} +
\begin{picture}(5.4,1)\thicklines\put(0.25,0.2){
        \put(0,0){\Grho}
        \put(3.5,0){\Grho}
        \put(2.5,0){\blue\oval(2,1.5)[b]}
        \put(2.5,0){\Red\oval(2,1.5)[t]}
        \put(2.4,0.75){\Red\vector(1,0){0.3}}
        \put(2.6,-0.75){\Blue\vector(-1,0){0.3}}
        \put(2.5,1.25){\makebox(0,0){\Red$ N^{-1}$}}
        \put(2.45,0.35){\makebox(0,0){\Blue$\pi$}}
        \put(2.5,-1.25){\makebox(0,0){\Blue N}}
        \put(3.5,0){\Blue\vector(-1,0){1.3}}
        \put(1.5,0){\Blue\line(1,0){1}}
        \put(0.05,0){
          \put(1.95,-1.1){\thinlines\line(1,2){0.3}}
          \put(2.35,-0.3){\thinlines\line(1,2){0.3}}
          \put(2.75,.5){\thinlines\line(1,2){0.3}}}
        }
\end{picture}\vphantom{\rule[-8mm]{0mm}{18mm}}}_
{\displaystyle\frac{
        {\Blue2m\Gamma_{\rho\;\pi^+\pi^-}} +
        {\Red2m\Gamma_{\rho\;\pi N N^{-1}}}}
        {\left(m^2-m_\rho^2-\mbox{Re}\Sigma^R\right)^2
        +{\Red m^2\Gamma_{\rm tot}^2}}} .
\end{eqnarray}
In principle, both diagrams have to be calculated in terms of fully
self-consistent propagators, i.e. with corresponding widths for all particles
involved. This formidable task has not been done yet. Using
micro-reversibility and the properties of thermal distributions, the two terms
in Eq. (\ref{A2}) contributing to the di-lepton yield (\ref{dndtdm}), can
indeed approximately be reformulated as the thermal average of a
$\pi^+\pi^-\rightarrow\rho\rightarrow{\rm e}^+{\rm e}^-$-annihilation process
and a $\pi N\rightarrow\rho N\rightarrow{\rm e}^+{\rm e}^-N$-scattering
process, i.e.
\begin{eqnarray}\label{x-sect}
\frac{\di n^{{\rm
  e}^+{\rm e}^-}}{\di m\di t}&\propto&
    \left<f_{\pi^+}f_{\pi^-}\; v_{\pi\pi}\;
      \sigma(\pi^+\pi^-\rightarrow\rho\rightarrow{\rm e}^+{\rm
          e}^-)\vphantom{A^A}\right. \cr&& +\left.
      f_{\pi}f_N\; v_{\pi N}\;\sigma(\pi N\rightarrow\rho N\rightarrow{\rm
          e}^+{\rm e}^-N)\vphantom{A^A}\right>_T ,
\end{eqnarray}
where $f_{\pi}$ and $f_N$ are corresponding particle occupations and
$v_{\pi\pi}$ and $v_{\pi N}$ are relative velocities.
However, the important fact to be noticed is that in order to preserve
unitarity the corresponding cross sections are no longer the free ones, as
given by the vacuum decay width in the denominator, but rather involve the
{\em medium dependent total width} (\ref{Gammatot}). This illustrates in
simple terms that rates of broad resonances can no longer simply be added in a
perturbative way.  Since it concerns a coupled channel problem, there is a
cross talk between the different channels to the extent that the common
resonance propagator attains the total width arising from all partial widths
feeding and depopulating the resonance. While a perturbative treatment with
free cross sections in Eq. (\ref{x-sect}) would enhance the yield at
resonance mass,
$m=m_{\rho}$, if a channel is added, cf. Fig.~2 left part, the correct
treatment (\ref{A2}) even inverts the trend and indeed depletes the yield at
resonance mass, right part in Fig.~2. Furthermore, one sees that only 
the total yield
involves the spectral function, while any partial cross section refers to
that partial term with the corresponding partial width in the numerator!
Unfortunately so far all these facts have been ignored or even overlooked in
the present transport treatment of broad resonances. 
\begin{figure}
\unitlength1cm
\begin{picture}(19,6.6)
\put(0,0){\epsfxsize=6cm\epsfbox{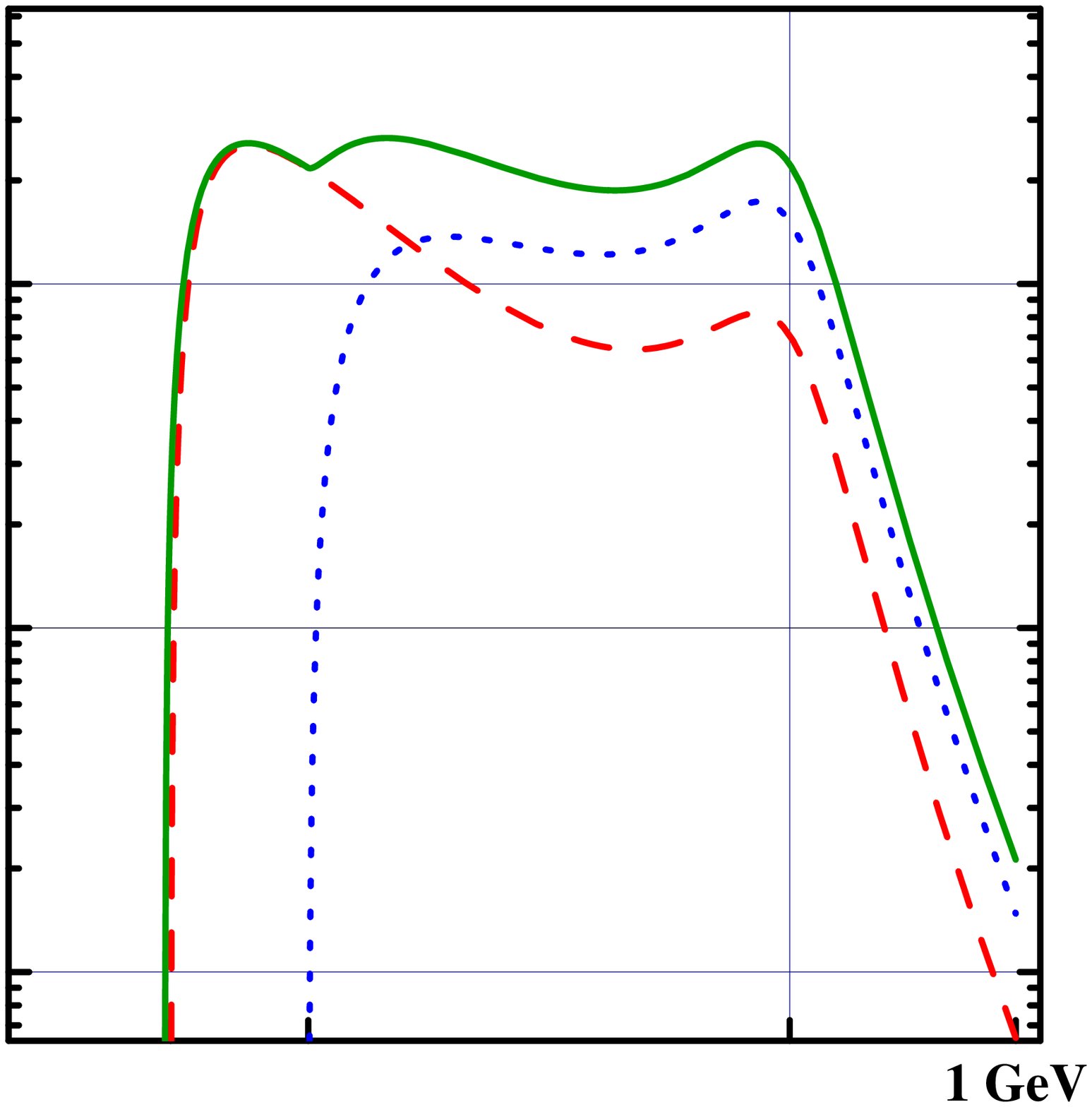}}
\put(6.,0){\epsfxsize=6cm\epsfbox{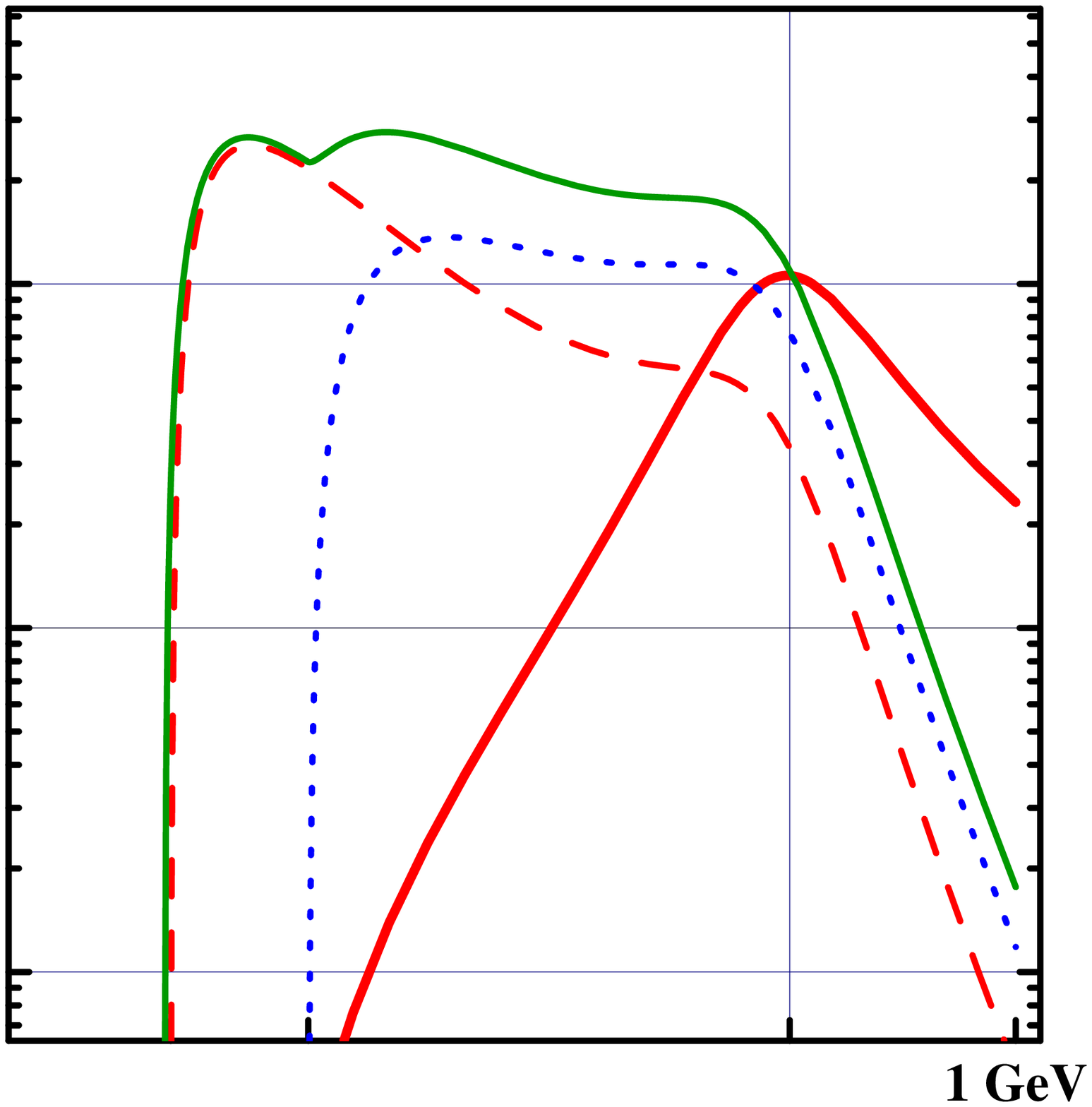}}
\put(6.,6.3){\makebox(0,0){\bf Di-lepton rates from thermal $\rho$-mesons
($T=110$ MeV)}}
\put(1.,0.1){\makebox(0,0){\small$m_{\pi}$}}
\put(1.8,0.1){\makebox(0,0){\small$2m_{\pi}$}}
\put(4.3,0.1){\makebox(0,0){\small$m_{\rho}$}}
\put(6,0){
\put(1.,0.1){\makebox(0,0){\small$m_{\pi}$}}
\put(1.8,0.1){\makebox(0,0){\small$2m_{\pi}$}}
\put(4.3,0.1){\makebox(0,0){\small$m_{\rho}$}}}
\put(3,5.6){\makebox(0,0){$\Gamma_{\rm tot}=\Gamma_{\rm free}$}}
\put(9,5.6){\makebox(0,0){full $\Gamma_{\rm tot}$}}
\put(9.5,3.0){\makebox(0,0){$A_{\rho}$}}
\end{picture}
\caption{
$\mbox{e}^+\mbox{e}^-$ rates (arb.  units)
  as a function of the invariant pair mass $m$ at $T=110$ MeV from
  $\pi^+\pi^-$ annihilation (dotted line) and direct $\rho$-meson contribution
  (dashed line), the full line gives the sum of both contributions. Left part:
  using the free cross section recipe, i.e. with $\Gamma_{\rm
    tot}=\Gamma_{\rho\;\pi^+\pi^-}$; right part: the correct partial rates
  (\ref{A2}). $A_{\rho}$ is given by the thick line. The calculations are done
  with $\Gamma_{\rho\leftrightarrow\pi\pi}(m_{\rho})=150$ MeV and
  $\Gamma_{\rho\leftrightarrow\pi N N^{-1}}(m_{\rho})=70$ MeV.}
\end{figure}
Compared to the spectral function both thermal components in Fig.~2 show a
significant enhancement on the low mass side and a strong depletion at high
masses due to the thermal weight $f\propto\exp(-p_0/T)$ in the rate
(\ref{dndtdm}).  
\unitlength=1mm
\begin{figure}
\begin{picture}(118,55)
\put(-10,74){{
\epsfig{file=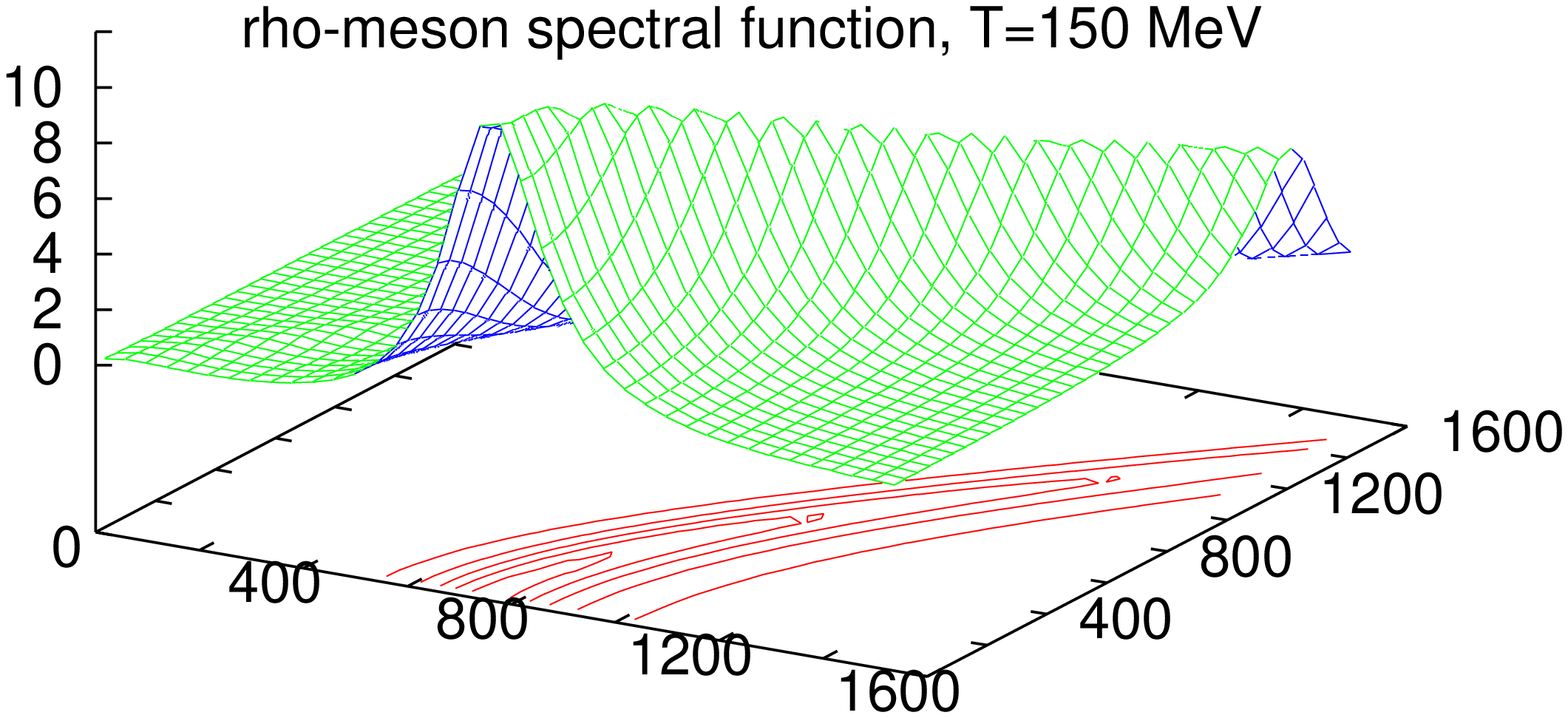,width=8cm,height=8cm,angle=-90}}}
\put(14,7){$\omega$}
\put(56,12){$\vec p$}
\put(60,55){{
\epsfig{file=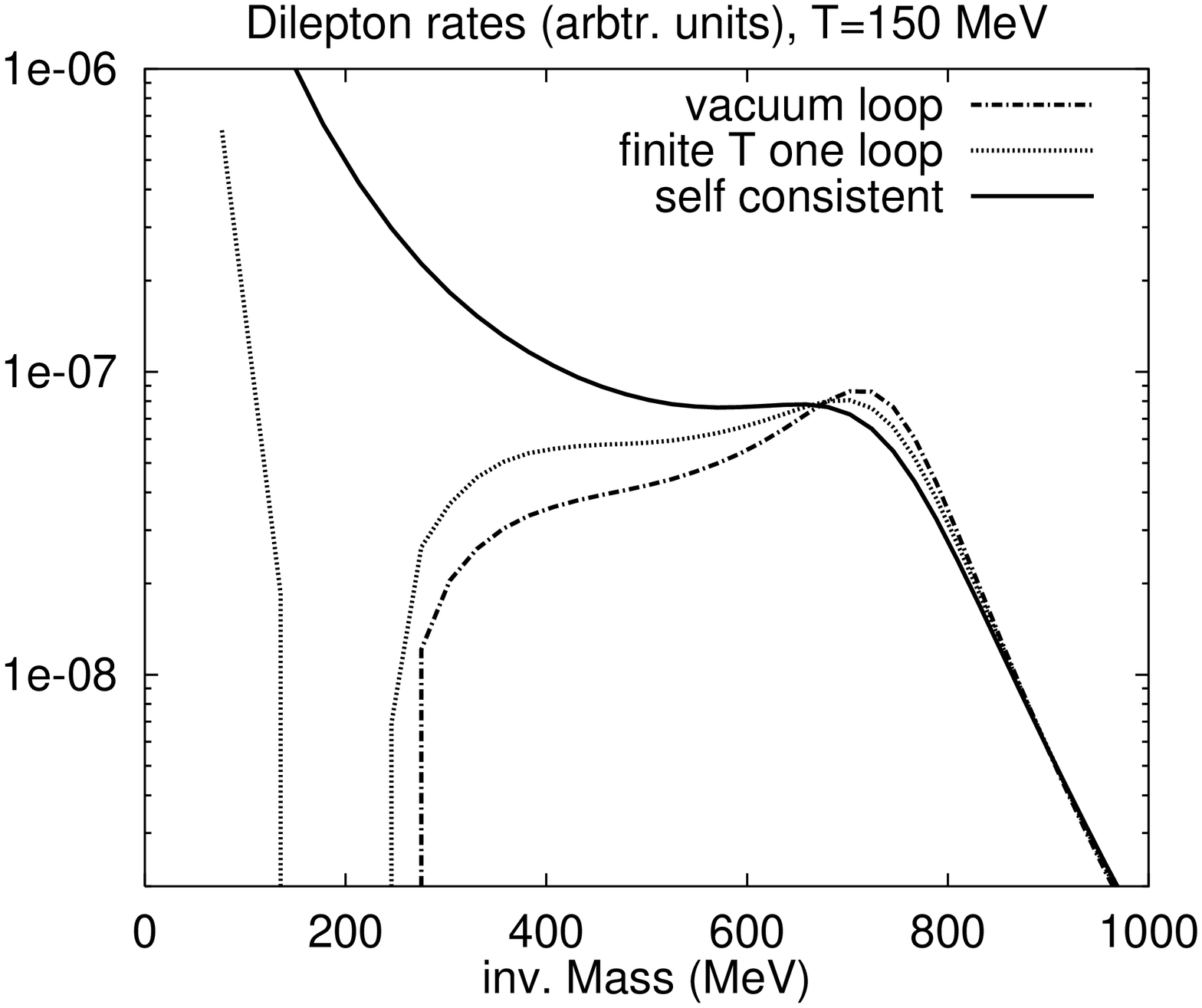,width=6cm,height=5.5cm,angle=-90}}}
\end{picture}
\caption{left part: contour plot of the self-consistent
  spectral function of the $\rho$-meson as a function of energy and spatial
  momentum; right part: thermal di-lepton rate as a function of invariant mass
  at ${\vec p}=300$ MeV$/c$}
\end{figure}

As an example we show an exploratory study of the interacting system of $\pi$,
$\rho$ and $a_1$-mesons described by the $\Phi$-functional 
\unitlength=1mm
\begin{fmffile}{kbd}
\fmfset{thin}{1.3pt}
\begin{eqnarray}\label{phi-pi-rho-a1}\cr
\Phi =\;
\parbox{12mm}{
\begin{fmfgraph*}(12,0)
\fmfpen{thin}
\fmfleft{l}
\fmfright{r}
\fmfforce{(0.0w,0.5h)}{l}
\fmfforce{(1.0w,0.5h)}{r}
\fmf{boson,label=$\rho$,label.side=left}{l,r}
\fmf{plain,left=.9,tension=.4,label=\raisebox{-.7mm}{$\pi$}}{l,r}
\fmf{plain,left=.9,tension=.4,label=\raisebox{-.7mm}{$\pi$},l.side=right}{r,l}
\fmfdot{l,r}
\end{fmfgraph*}
}
\;+\;
\parbox{12mm}{
\begin{fmfgraph*}(12,0)
\fmfpen{thin}
\fmfleft{l}
\fmfright{r}
\fmfforce{(0.0w,0.5h)}{l}
\fmfforce{(1.0w,0.5h)}{r}
\fmf{plain,left=.9,tension=.4,label=\raisebox{-.7mm}{$\pi$}}{l,r}
\fmf{boson,label=$\rho$,l.side=left}{l,r}
\fmf{gluon,left=.9,tension=.4,label=\raisebox{1.5mm}{$a_1$},l.side=right}{r,l}
\fmfdot{l,r}
\end{fmfgraph*}
}
\;+\;
\parbox{12mm}{
\begin{fmfgraph*}(12,0)
\fmfpen{thin}
\fmfleft{l}
\fmfright{r}
\fmfforce{(0.0w,0.5h)}{l}
\fmfforce{(1.0w,0.5h)}{r}
\fmf{plain,left=.9,tension=.4,label=\raisebox{-.7mm}{$\pi$},l.side=right}{r,l}
\fmf{plain,left=.3,tension=.4,label=\raisebox{-.7mm}{$\pi$},l.side=right}{r,l}
\fmf{plain,left=.3,tension=.4,label=\raisebox{-.7mm}{$\pi$}}{l,r}
\fmf{plain,left=.9,tension=.4,label=\raisebox{-.7mm}{$\pi$}}{l,r}
\fmfdot{l,r}
\end{fmfgraph*}
}\vphantom{\rule[-4mm]{0mm}{4mm}}
\end{eqnarray}
\end{fmffile}
\noindent (cf.  section
\ref{Phi-derv} below). The couplings and masses are chosen as to reproduce the
known vacuum properties of the $\rho$ and $a_1$ meson with nominal masses and
widths $m_{\rho}=770$ MeV, $m_{a_1}=1200$ MeV, $\Gamma_{\rho}=150$ MeV,
$\Gamma_{a_1}=400$ MeV. The results of a finite temperature calculation at
$T=150$ MeV with all self-energy loops resulting from the $\Phi$-functional of
Eq. (\ref{phi-pi-rho-a1}) computed~\cite{Hees} with self-consistent broad
width Green's 
functions are displayed in Fig. 3 (corrections to the real part of the
self-energies were not yet included). The last diagram of $\Phi$ with the four
pion self-coupling has been added in order to supply pion with broad
mass-width as they would result from the coupling of pions to nucleons and the
$\Delta$ resonance in nuclear matter environment. As compared to first-order
one-loop results which drop to zero below the 2-pion threshold at 280 MeV, the
self-consistent results essentially add in strength at the low-mass side of
the di-lepton spectrum.\\[-6mm]
\section{Quantum Kinetic Equation}

The two above-presented examples unambiguously show that for
consistent dynamical treatment of nonequilibrium evolution of soft
radiation and broad resonances we need a transport theory that takes
due account of mass-widths of constituent particles. A proper frame
for such a transport is provided by Kadanoff--Baym equations. We
consider the Kadanoff--Baym equations 
in the first-order gradient approximation,
assuming that time--space evolution of a system is smooth enough to
justify this approximation. 

First of all, it is helpful to avoid all the imaginary factors inherent in the
standard Green function formulation ($\Gr^{ij}$ with $i,j\in\{-+\}$) and
introduce quantities which are real and, in the quasi-homogeneous limit,
positive and therefore have a straightforward physical interpretation, much
like for the Boltzmann equation.  In the Wigner representation we define
%
\begin{eqnarray}
\label{F}
\Fd (X,p) &=& \A (X,p) \fd (X,p)
 =  (\mp )\ii \Gr^{-+} (X,p) , \nonumber\\
\Fdt (X,p) &=& \A (X,p) [1 \mp \fd (X,p)] = \ii \Gr^{+-} (X,p), \\ 
\label{A}
 A (X,p) &\equiv& -2\Im \Gr^R (X,p) = \Fdt \pm \Fd =
\ii \left(\Ga^{+-}-\Ga^{-+}\right)
\end{eqnarray}
%
for the generalized Wigner functions $\F$ and $\Ft$ with the corresponding
{\em four}-phase-space distribution functions $\fd(X,p)$ and Fermi/Bose
factors $[1 \mp \fd (X,p)]$, with the spectral function $A(X,p)$ and the
retarded 
propagator $\Gr^R$.  Here and below the upper sign corresponds to fermions and
the lower one, to bosons. According to relations between Green functions
$\Gr^{ij}$ {\em only two independent real functions of all the $\Gr^{ij}$ are
  required for a complete description}.  Likewise the reduced gain and loss
rates of the collision integral and the damping rate are defined as
%
\def\ga{\gamma}
\begin{eqnarray}
\label{gain}
\Ldt (X,p) 
&=&   (\mp )\ii \Se^{-+} (X,p), \quad
\Ld (X,p)  
=  \ii \Se^{+-} (X,p), \\
\label{G-def}
\Gamma (X,p)&\equiv& -2\Im \Se^R (X,p) = \Ld (X,p)\pm\Ldt (X,p), 
\end{eqnarray}
%
where $\Se^{ij}$  are contour components of the self-energy, and
$\Se^R$ is the retarded self-energy. 
 
In terms of this notation and within the
first-order gradient approximation, the Kadanoff--Baym equations 
for $\Fd$ and $\Fdt$ (which result from differences of the 
corresponding Dyson's equations with their adjoint ones) take the 
kinetic form 
%
\begin{eqnarray} 
\label{keqk1} 
{\cal D}\Fd  -  
\Pbr{\Ldt , 
\Re\Gr^R} &=& C  , 
\\
\label{keqkt1} 
{\cal D}\Fdt - 
\Pbr{\Ld , \Re\Gr^R} &=&\mp C  
\end{eqnarray} 
%
with 
the drift operator and collision term respectively
%
\begin{eqnarray}
\label{Coll(kin)}
{\cal D}&=&\left(
\vu_{\mu} - 
\frac{\partial \Re\Sa^R}{\partial p^{\mu}} 
\right) 
\partial^{\mu}_X + 
\frac{\partial \Re\Sa^R}{\partial X^{\mu}}  
\frac{\partial }{\partial p_{\mu}}, 
\quad\quad v^\mu=(1,\vec{p}/m), 
\\
C (X,p) &=&
\Ldt (X,p) \Ft (X,p) 
- \Ld (X,p) \F (X,p).
\end{eqnarray}
%

Within the same approximation level there are two
alternative equations for $\Fd$ and $\Fdt$
%
\begin{eqnarray}
\label{mseq(k)1}
M\Fd - \Re\Ga^R\Ldt
&=&\frac{1}{4}\left(\Pbr{\Gm,\Fd} - \Pbr{\Ldt,\A}\right),
\\\label{mseqt(k)1}
M\Fdt - \Re\Ga^R\Ld
&=&\frac{1}{4}\left(\Pbr{\Gm,\Fdt} - \Pbr{\Ld,\A}\right)
\end{eqnarray}
%
with the  ``mass'' function 
$M(X,p)=p_0 -\vec{p}^2/2m -\Re\Se^R (X,p)$. 
These two equations result from sums of the corresponding Dyson's equations 
with their adjoint ones. Eqs. (\ref{mseq(k)1}) and (\ref{mseqt(k)1})
can be called the mass-shell equations, 
since in the quasiparticle limit they provide the on-mass-shell condition
$M=0$. 
Appropriate combinations of the two sets
(\ref{keqk1})--(\ref{keqkt1}) and (\ref{mseq(k)1})--(\ref{mseqt(k)1}) provide
us with retarded Green's function equations, 
which are simultaneously solved~\cite{Kad62,BM} by
%
\begin{eqnarray}
\label{Asol}
\Gr^R=\frac{1}{M(X,p)+\ii\Gamma(X,p)/2}\Rightarrow
\left\{\begin{array}{rcl}
A &=&\displaystyle
\frac{\Gamma}{M^2 + \Gamma^2 /4},\\[2mm]
\Re\Gr^R &=& \displaystyle 
\frac{M}{M^2 + \Gamma^2 /4}.
\end{array}\right. 
\end{eqnarray}
%

With the solution (\ref{Asol}) for $\Gr^R$ equations (\ref{keqk1}) and
(\ref{keqkt1}) become identical to each other, as well as Eqs.
(\ref{mseq(k)1}) and (\ref{mseqt(k)1}). However, Eqs.
(\ref{keqk1})--(\ref{keqkt1}) still are not identical to Eqs.
(\ref{mseq(k)1})--(\ref{mseqt(k)1}), while they were identical before the
gradient expansion. Indeed, one can show~\cite{IKV99} that Eqs.
(\ref{keqk1})--(\ref{keqkt1}) differ from Eqs.
(\ref{mseq(k)1})--(\ref{mseqt(k)1}) in second order gradient terms. This is 
acceptable within the gradient approximation, however, the still remaining
difference in the second-order terms is inconvenient from the practical point
of view.  Following Botermans and Malfliet~\cite{BM}, we express $\Ldt=\Gm
f+O(\partial_X)$ and $\Ldt=\Gm (1\mp f)+O(\partial_X)$ from the l.h.s. of
mass-shell Eqs.  (\ref{mseq(k)1}) and (\ref{mseqt(k)1}), substitute them into
the Poisson bracketed terms of Eqs. (\ref{keqk1}) and (\ref{keqkt1}), and
neglect all the resulting second-order gradient terms.  The so obtained {\em
  quantum four-phase-space kinetic equations for $\Fd=fA$ and $\Fdt=(1\mp f)
  A$} then read
%
\begin{eqnarray}
\label{keqk}
\Do 
\left(f A\right) - 
\Pbr{\Gm f,\Re\Gr^R} &=& C , 
\\
\label{keqkt}
\Do 
\left((1\mp f) A\right) - 
\Pbr{\Gm (1\mp f),\Re\Gr^R} &=& \mp C .  
\end{eqnarray}
%
These quantum four-phase-space kinetic equations, which are identical to each
other in view of retarded relation (\ref{Asol}), are at the same time
completely identical to the correspondingly substituted mass-shell Eqs.
(\ref{mseq(k)1}) and (\ref{mseqt(k)1}).

The validity of the gradient approximation~\cite{IKV99} relies on the overall
smallness of the collision term $C=\{\mbox{gain} - \mbox{loss}\}$ rather than
on the smallness of the damping width $\Gamma$. Indeed, while fluctuations and
correlations are governed by time scales given by $\Gamma$, the Kadanoff--Baym
equations describe the behavior of the ensemble mean of the occupation in
phase-space $F(X,p)$. It implies that $F(X,p)$ varies on space-time scales
determined by $C$. In cases where $\Gamma$ is not small enough by itself, the
system has to be sufficiently close to equilibrium in order to provide a valid
gradient approximation through the smallness of the collision term $C$.
Both the Kadanoff--Baym (\ref{keqk1}) and the Botermans--Malfliet choice
(\ref{keqk}) are, of course, equivalent within the validity range of the
first-order gradient approximation. Frequently, however, such equations are
used beyond the limits of their validity as ad-hoc equations, and then the
different 
versions may lead to different results. So far we have no physical condition
to prefere one of the choices. The procedure,
where in all Poisson brackets the $\Ldt$ and $\Ld$ terms have consistently
been replaced by $f\Gamma$ and $(1\mp f)\Gamma$, respectively, is therefore
optional. However, in doing so we gained some advantages. Beside the fact that
quantum four-phase-space kinetic equation (\ref{keqk}) and the mass-shell
equation are then {\em exactly} equivalent to each other, this set of
equations has a particular features with respect to the definition of a
nonequilibrium entropy flow in connection with the formulation of an {\em
  exact} H-theorem in certain cases.  If we omit these substitutions, both
these features would become approximate with deviations at the second-order
gradient level.


The equations so far presented, mostly with the Kadanoff--Baym choice
(\ref{keqk1}), were the starting point for many derivations of
extended Boltzmann and generalized kinetic equations, ever since these
equations have been formulated in 1962.  Most of those derivations use
the equal-time reduction by integrating the four-phase-space equations
over energy $p_0$, thus reducing the description to three-phase-space
information, cf. 
refs.~\cite{Bez,LSV,LipS,Kraft,Bonitz,Bornath,SCFNW,Jeon,VBRS} and
refs.  therein. This can only consistently be done in the limit of small width
$\Gamma$ employing some kind of quasi-particle ansatz for the spectral
function $A(X,p)$. Particular attention has been payed to the treatment of the
time-derivative parts in the Poisson brackets, which in the four-phase-space
formulation still appear time-local, i.e. Markovian, while they lead to
retardation effects in the equal-time reduction. Generalized quasiparticle
ans\"atze were proposed, which essentially improve the quality and consistency
of the approximation, providing those extra terms to the naive Boltzmann
equation (some times called additional collision term) which are responsible
for the correct second-order virial corrections and the appropriate
conservation of total energy, cf. \cite{LipS,Bornath} and refs. therein.
However, all these derivations imply some information loss about the
differential mass spectrum due to the inherent reduction to a 3-momentum
representation of the distribution functions by some specific ansatz.  With
the aim to treat cases as those displayed in Figs. 2 and 3, where the
differential mass spectrum can be observed by di-lepton spectra, within a
self-consistent non-equiblibrium approach, one has to treat the differential
mass information dynamically, i.e. by means of Eq. (\ref{Asol}) avoiding any
kind of quasi-particle reductions and work with the full quantum
four phase-space kinetic Eq.  (\ref{keqk}). In the following we discuss the
properties of this set of quantum kinetic equations.


\section{$\Phi$-derivable approximations}\label{Phi-derv}

The preceding considerations have shown that one needs a transport scheme
adapted to broad resonances. Besides the conservation laws it should comply
with requirements of unitarity and detailed balance. A practical suggestion
has been given in ref.~\cite{DB} in terms of cross-sections. However, this
picture is tied to the concept of asymptotic states and therefore not well
suited for the general case, in particular, if more than one channel feeds
into a broad resonance. Therefore, we suggest to revive the so-called
$\Phi$-derivable scheme, originally proposed by Baym~\cite{Baym} on the basis
of the generating functional, or partition sum, given by Luttinger and
Ward~\cite{Luttinger}, and later reformulated in terms of
path-integrals~\cite{Cornwall}.  The auxiliary functional $\Phi$ is given by
two-particle irreducible vacuum diagrams. It solely depends on fully
re-summed, i.e. self-consistently generated propagators $\ii \Gr(x,y)
=< T_{C}\widehat{\varphi}(x)\widehat{\varphi}^{\dagger}(y)>$,
where $T_{C}$ indicates real-time contour ordering. The
consistency is provided by the fact that $\Phi$ is the generating functional
for the re-summed self-energy $\Se(x,y)$ via functional variation of $\Phi$
with respect to any contour ordered propagator $\Gr(y,x)$, i.e.
%
\begin{eqnarray}\label{varphi}
-\ii 
\Se (x,y) =\mp \delta \ii \Phi / \delta \ii \Gr(y,x).  
\end{eqnarray}
%
%
%
An extension to include classical fields or condensates into the scheme is
presented in ref.~\cite{IKV} In graphical terms, the variation (\ref{varphi})
with respect to $\Gr$ is realized by opening a propagator line in all diagrams
of $\Phi$.  The resulting set of thus opened diagrams must then be that of
proper skeleton diagrams of $\Se$ in terms of {\em full propagators}, i.e.
void of any self-energy insertions. As a consequence, the $\Phi$-diagrams have
to be {\em two-particle irreducible}, i.e. they cannot be decomposed into two
pieces by cutting two propagator lines.

The key property is that truncating the auxiliary functional $\Phi$ to a limited
subset of diagrams leads to a self-consistent, i.e closed, approximation
scheme. Thereby the approximate forms of $\Phi$ define {\em
  effective} theories, where $\Phi^{\scr{(appr.)}}$ serves as a generating
functional for the approximate self-energies $\Sa^{\scr{(appr.)}}(x,y)$
through relation (\ref{varphi}), which then enter as driving terms for the
Dyson's equations of the different species in the system.  As Baym~\cite{Baym}
has shown, such a $\Phi$-derivable approximation is conserving as
related to global symmetries of the original theory. 
We explicitly cite the forms of the conserved Noether current
and of the energy--momentum tensor, cf. ref.~\cite{IKV},
%
\begin{eqnarray}
\label{c-new-currentk}
j^{\mu} (X) 
&=&\frac{e}{2} \int \dpi{p}
\vu^{\mu} 
\left(\Fd (X,p) \mp \Fdt (X,p) \right),\hspace*{-1cm} \\
\label{E-M-new-tensork}
\Theta^{\mu\nu}(X)
&=&\frac{1}{2} \int \dpi{p} 
\vu^{\mu} p^{\nu} 
\left(\Fd (X,p) \mp \Fdt (X,p) \right)
+ g^{\mu\nu}\left(
{\cal E}^{\scr{int}}-{\cal E}^{\scr{pot}}
\right),
\end{eqnarray}
%
where 
%
\begin{eqnarray}
\label{eps-int} 
{\cal E}^{\scr{int}}(X)=\left<-\Lint(X)\right>
=-\left.\frac{\delta\Phi}{\delta\lambda(X)}\right|_{\lambda=1}\nonumber
\end{eqnarray}
%
is the density of interaction energy ($\lambda(X)$ locally
scales the coupling strength of vertices, cf. ref.~\cite{IKV}) and 
the density of potential energy ${\cal E}^{\scr{pot}}$ takes the
following simple form within the first-order gradient approximation
%
\begin{eqnarray}
\label{eps-potk}
{\cal E}^{\scr{pot}}(X)
= \frac{1}{2}
\int\dpi{p} \left[
\Re\Sa^R \left(\Fd\mp\Fdt\right)
+ \Re\Ga^R\left(\Gb\mp\Gbt\right)\right].
\nonumber
\end{eqnarray}
%
The first term of ${\cal E}^{\scr{pot}}$ complies with
quasi-particle expectations, namely mean potential times density, the second
term displays the role of fluctuations $I=\Gb\mp\Gbt$ in the potential energy
density.  This fluctuation term precisely arises form the Poisson bracket term
in the kinetic Eq. (\ref{keqk}) which induces a back-flow. It restores the
Noether expressions (\ref{c-new-currentk}) and (\ref{E-M-new-tensork})
as being indeed the exactly conserved
quantities. In this compensation we see the essential role of the fluctuation
term in the quantum four-phase-space kinetic equation. Dropping or
approximating this term 
would spoil the conservation laws. Before the gradient expansion, quantities
(\ref{c-new-currentk}) and (\ref{E-M-new-tensork}) are {\em exact}
integrals of equations of motion. While after the gradient expansion, 
they comply with the quantum four-phase-space kinetic equation 
(\ref{keqk}) up to the first-order gradient terms.  

At the same time the $\Phi$-derivable scheme provides thermodynamical
consistency. The latter automatically implies correct detailed balance
relations between the various transport processes. For multicomponent systems
it leads to a {\em actio} = {\em reactio} principle. This implies that the
properties of one species are not changed by the interaction with other
species without affecting the properties of the latter ones, too.  Some
thermodynamic examples have been considered recently, e.g., for the
interacting $\pi N \Delta$ system~\cite{Weinhold} and for a relativistic QED
plasma~\cite{Vanderheyden}.

\section{Collision Term}

To further discuss the transport treatment we need an explicit form of the
collision term (\ref{Coll(kin)}), which is provided from the $\Phi$ functional
in the $-+$ matrix notation via the variation rules (\ref{varphi})
as
%
\def\tp{p'}\def\tm{m'}\def\tW{\widetilde{W}} 
\def\tR{\widetilde{R}}
\begin{eqnarray}
\label{Coll-var} 
C (X,p) =&& 
\frac{\delta\ii\Phi}{\delta\Ft(X,p)}\Ft(X,p)
-\frac{\delta\ii\Phi}{\delta\F(X,p)}\F(X,p) .
\end{eqnarray}
Here we assumed $\Phi$ be transformed into the Wigner representation and all
$\mp\ii\Gr^{-+}$ and $\ii\Gr^{+-}$ to be replaced by the Wigner-densities
$\Fd$ and $\Fdt$. Thus, the structure of the collision term can be inferred
from the structure of the diagrams contributing to the functional $\Phi$. To
this end, in close analogy to the consideration of ref.~\cite{KV}, we
discuss various decompositions of the $\Phi$-functional, from which the in-
and out-rates are derived. For the sake of physical transparency, 
we confine our treatment to the {\em local} case, where in the
Wigner representation all the Green functions are taken at the 
same space-time coordinate $X$ and all non-localities, i.e. 
derivative corrections, are disregarded.  Derivative corrections 
give rise to memory effects in the collision term, which will be 
analyzed separately for the specific case of the triangle diagram.

Consider a given closed diagram of $\Phi$, at this level specified by a
certain number $n_{\lambda}$ of vertices and a certain contraction pattern.
This fixes the topology of such a contour diagram. It leads to $2^{n_\lambda}$
different diagrams in the $-+$ notation from the summation over all $-+$ signs
attached to each vertex.  Any $-+$ notation diagram 
of $\Phi$, which contains vertices of either sign, can
be decomposed into two pieces
in such a way that each of the
two sub-pieces contains vertices of only one type of sign\footnote{To
  construct the decomposition, just deform a given mixed-vertex diagram of
  $\Phi$ in such a way that all $+$ and $-$ vertices are placed left and
  respectively right from a vertical division line and then cut along this
  line.}
%
\begin{eqnarray}\label{decomp}\unitlength6mm
&&\ii \Phi_{\alpha\beta}=\unitlength7mm
\begin{picture}(4.5,1.5)
\put(0.,.1){\phidecomposition}
\end{picture}\vphantom{\sum_{\int}}
    =\left(\alpha\left|
\Fd_1 ...  \Fdt'_{1} ...\right|\beta\right)
\\
    &\Rightarrow&
    \int\frac{\di^4 p_1}{(2\pi)^4}\cdots\frac{\di^4 p'_1}{(2\pi)^4}\cdots
    (2\pi)^4
    \delta^4\left(\sum_{i} p_i - \sum_{i} p'_i \right)
    V^*_{\alpha}
    \Fd_1 ... \Fdt'_{1} ...
    V_{\beta}
\nonumber
\end{eqnarray}
%
with $\Fd_1\cdots\Fd_m\Fdt'_1\cdots\Fdt'_{\tilde m}$ linking the two 
amplitudes.  The $V^*_{\alpha}(X;p_1,...p'_1,...)$ and 
$V_{\beta}(X;p_1,...p'_1,...)$ amplitudes represent multi-point vertex 
functions of only one sign for the vertices, i.e.  they are either entirely 
time ordered ($-$ vertices) or entirely anti-time ordered ($+$ vertices). 
Here we used the fact that adjoint expressions are complex conjugate to 
each other.  Each such vertex function is determined by normal Feynman 
diagram rules.  Applying the matrix variation rules (\ref{Coll-var}), we 
find that the considered $\Phi$ diagram gives the following contribution to 
the local part of the collision term (\ref{Coll(kin)}) 
%
\def\tp{p'}\def\tm{m'}\def\tW{\widetilde{W}} 
\def\tR{\widetilde{R}}
\begin{eqnarray}
\label{Coll-var-loc} 
\hspace*{-5mm}
&&C^{\scr{loc}} (X,p) \Rightarrow \frac{1}{2}
 \int \dpi{p_1} \cdots \dpi{\tp_1}\cdots 
R 
\left[\sum_{i} \delta^4(p_i-p)
-\sum_{i} \delta^4(\tp_i-p)\right]
\nonumber\\
&&\times 
\left\{
\Ft_1... \F'_1...
-
\F_1... \Ft'_1... 
\right\}
(2\pi)^4\delta^4\left(\sum_{i} p_i - \sum_{i} \tp_i \right).
\label{Multi-rate}
\end{eqnarray}
%
with the partial process rates
%
\begin{equation}\label{Rmm}
R(X;p_1,...p'_1,...)
=
\sum_{(\alpha\beta)\in \Phi} 
\Re\left\{
V^*_{\alpha}(X;p_1,...p'_1,...)
V_{\beta}(X;p_1,...p'_1,...)\right\}.
\end{equation}
%
The restriction to the real part arises, since with $(\alpha|\beta)$ also the
adjoint $(\beta|\alpha)$ diagram contributes to this collision term.
However these rates are not necessarily positive. In this
point, the generalized scheme differs from the conventional Boltzmann
kinetics. 

An important example of approximate $\Phi$ which we 
extensively use below is 
%
\begin{eqnarray}
\label{Phi-ring}& &\nonumber\\[-11mm] 
\ii \Phi & &=  \frac{1}{2} \loopa + \frac{1}{4} \loopb{ } 
+ \frac{1}{6} \loopc{ }  \vphantom{\rule[-6mm]{0mm}{4mm}}
\end{eqnarray}
%
where logarithmic factors due to the special features of the
$\Phi$-diagrammatic technique are written out explicitly,
cf. ref.~\cite{IKV99}. 
In this example we assume a system of fermions interacting via a two-body
potential $V=V_0 \delta(x-y)$, and, for the sake of simplicity, disregard its
spin structure. The $\Phi$ functional of Eq. (\ref{Phi-ring}) results in the 
following local collision term 
%
\begin{eqnarray}
\label{C30} 
C^{\scr{loc}} 
&=& d^2
\int \dpi{p_1} \dpi{p_2} \dpi{p_3}
\left(
\left|\;\; \wa{-} \;\;+ \!\!\!\!\!\wb{-} \;\;\right|^2 - 
\left|\!\!\!\!\!\wb- \;\;\right|^2\right)
\nonumber 
\\[1mm]
&&\times 
\delta^4\left(p + p_1 - p_2 - p_3\right) 
\left(
\F_2\F_3 \Ft\Ft_1 -
\Ft_2\Ft_3 \F\F_1
\right),  
\end{eqnarray}
%
where $d$ is the spin (and maybe isospin) degeneracy factor. From this 
example one can see that the positive definiteness of transition rate is 
not evident. 

The first-order gradient corrections to the local collision term
(\ref{Multi-rate}) are called {\em memory} corrections. 
{\em Only diagrams of third and
higher order in the number of vertices give rise to memory effects}. 
In particular, only the last diagram of Eq. (\ref{Phi-ring}) gives rise to 
the memory correction, which is calculated in ref.~\cite{IKV99}.

\section{Entropy} \label{Entropy}

Compared to exact descriptions, which are time reversible, reduced description
schemes in terms of relevant degrees of freedom have access only to some
limited information and thus normally lead to irreversibility. 
In the Green's function formalism presented here the information loss
arises from the truncation of the exact Martin--Schwinger hierarchy, where the
exact one-particle Green function couples to the two-particle Green functions,
cf.  refs.~\cite{Kad62,BM}, which in turn are coupled to the three-particle
level, etc. This
truncation is achieved by the standard Wick decomposition, where all
observables are expressed through one-particle propagators and therefore
higher-order correlations are dropped. This step provides the Dyson's equation
and the corresponding loss of information is expected to lead to
a growth of entropy with time.  

We start with general manipulations which lead us to definition of the kinetic
entropy flow.  We multiply Eq. (\ref{keqk}) by $-\ln(\F/A)$, 
Eq. (\ref{keqkt}) by 
$(\mp)\ln(\Ft/A)$, take their sum, integrate it over $\di^4
p/(2\pi )^4$, and finally sum the result over internal degrees of freedom like
spin ($\Tr$). Then we arrive at the following relation
%
\begin{eqnarray}
\label{s-Eq.} 
\partial_\mu s^\mu_{\scr{loc}} (X) =\mbox{Tr}
\int \dpi{p} \ln \frac{\Ft_a}{\F} C (X,p),   
\end{eqnarray}
%
where the quantity 
\begin{eqnarray}\label{entr-transp}
s^\mu_{\scr{loc}}  = \mbox{Tr}\int \dpi{p} 
\frac{A^2 \Gamma }{2}
\left[\left(
\vu^{\mu} - \frac{\partial \Re\Se^R }{\partial p_\mu}
\right)
- M \Gamma^{-1}\frac{\partial \Gamma}{\partial p_\mu}
\right] 
\sigma (X,p)
\end{eqnarray}
(where $\sigma (X,p)=\mp [1\mp f]\ln [1\mp f]-f\ln f$)
obtained from the l.h.s. of the kinetic equation is interpreted as the local
(Markovian) part of the entropy flow. Indeed, the $s^0_{\scr{loc}}$ has 
proper thermodynamic and quasiparticle limits~\cite{IKV99}. However, to be 
sure that this is indeed the entropy flow we must prove the H-theorem for 
this quantity. 

First, let us consider the case, when memory corrections to the collision 
term are negligible. Then we can make use of the form
(\ref{Multi-rate}) of the local 
collision term. Thus, we arrive at the relation
%
\begin{eqnarray}
\label{s(coll)} 
&&\mbox{Tr}\int \dpi{p} \ln \frac{\Ft}{\F} C_{\scr{loc}} (X,p) 
\Rightarrow \mbox{Tr}
\frac{1}{2} 
 \int \dpi{p_1}\cdots \dpi{\tp_1}\cdots
R 
\nonumber 
\\
&&\times
\left\{\F_1... \Ft'_1 ... 
-
\Ft_1 ... \F'_1 ...\right\}
\ln\frac{\F_1 ... \Ft'_1...}
        {\Ft_1...\F'_1 ...}
(2\pi)^4
\delta^4\left(\sum_{i} p_i - \sum_{i} \tp_i \right).
\end{eqnarray}
%
In case all rates $R$ are
non-negative, i.e. $R\ge 0$, this expression is non-negative, since
$(x-y)\ln(x/y) \ge 0$ for any positive $x$ and $y$.  In particular,
$R\ge 0$ takes place for all $\Phi$-functionals up to two vertices.
Then the divergence of $s_{\scr{loc}}^\mu$ is non-negative 
%
\begin{eqnarray}\label{dmusmu>0}
\partial_\mu s^\mu_{\scr{loc}}(X)\ge 0,
\end{eqnarray}
%
which proves the $H$-theorem in this case with (\ref{entr-transp}) as the 
nonequilibrium entropy flow.
However, as has been mentioned above, we are unable to show that $R$
always takes non-negative values for all $\Phi$-functionals. 

If memory corrections are essential, the situation is even more involved. Let
us consider this situation again at the example of the $\Phi$ approximation 
given by Eq. (\ref{Phi-ring}). 
We assume that the fermion--fermion potential interaction is such that the
corresponding transition rate of the corresponding local 
collision term (\ref{C30}) is always non-negative, so that the $H$-theorem 
takes place in the local approximation, i.e. when we keep only 
$C^{\scr{loc}}$. 
Here we will schematically describe calculations of 
ref.~\cite{IKV99} which, to our opinion, 
illustrate a general strategy for the derivation of memory correction to the
entropy, provided the $H$-theorem holds for the local part.

Now Eq. (\ref{s-Eq.}) takes the form
%
\begin{equation}
\label{log-term}
\partial_\mu s^\mu_{\scr{loc}} (X) =\mbox{Tr}
\int \dpi{p} \ln \frac{\Ft}{\F} C^{\scr{loc}}
+\mbox{Tr}\int \dpi{p} \ln \frac{\Ft}{\F} C^{\scr{mem}},    
\end{equation}
%
where $ s^\mu_{\scr{loc}}$ is still the Markovian entropy flow defined by 
Eq.  (\ref{entr-transp}). Our aim here is to present the last term on the 
r.h.s. of Eq.  (\ref{log-term}) in the form of full $x$-derivative 
%
\begin{equation}
\label{ent-mar}
\mbox{Tr}\int \dpi{p} \ln \frac{\Ft}{\F} C^{\scr{mem}} 
=-\partial_{\mu} s_{\scr{mem}}^{\mu} (X) + \delta c_{\scr{mem}} (X)
\end{equation}
%
of some function $s_{\scr{mem}}^{\mu} (X)$, which we then interpret as a
non-Markovian correction to the entropy flow of Eq. (\ref{entr-transp}) 
plus a
correction ($\delta c_{\scr{mem}}$).
For the memory induced by the triangle diagram of Eq.(\ref{Phi-ring})
detailed calculations of ref.~\cite{IKV99} show that 
the smallness of the $\delta 
c_{\scr{mem}}$, originating from small space--time gradients and small 
deviation from equilibrium, allows us to neglect this term as compared to 
the first term in r.h.s. of Eq. (\ref{ent-mar}).
Thus, we obtain
%
\begin{equation}
\label{H-der}
\partial_\mu \left(s^\mu_{\scr{loc}} + s_{\scr{mem}}^{\mu}\right) \simeq
\mbox{Tr}\int \dpi{p} \ln \frac{\Ft}{\F} C^{\scr{loc}} \geq 0, 
\end{equation}
%
which is the $H$-theorem for the non-Markovian kinetic equation under
consideration with $s^\mu_{\scr{loc}} + s_{\scr{mem}}^{\mu}$ as the proper
entropy flow. The r.h.s. of Eq. (\ref{H-der}) is non-negative 
provided the corresponding transition rate in the local 
collision term of Eq. (\ref{C30}) is non-negative.

The explicit form of $s_{\scr{mem}}^{\mu}$ is very complicated, see 
ref.~\cite{IKV99}. In equilibrium 
at low temperatures we get $s_{\scr{mem}}^0
\sim T^3 \ln T$ which gives the leading correction to the standard Fermi-liquid
entropy.  This is the famous correction~\cite{Baym91,Carneiro} 
to the specific heat of
liquid $^3$He. Since this correction is quite
comparable (numerically) to the leading term in the specific heat ($\sim T$),
one may claim that {\em liquid $^3$He is a liquid with quite strong memory
effects from the point of view of kinetics}.

\section{Summary}

A number of problems arising in different dynamical systems, e.g. in
heavy-ion collisions, require an explicit
treatment of dynamical evolution of particles with finite mass-width. This was
demonstrated for the example of bremsstrahlung from a nuclear source, where
the soft part of the spectrum can be reproduced only provided the mass-widths
of nucleons in the source are taken explicitly into account. In this case the
mass-width arises due to collisional broadening of nucleons. Another example
considered concerns propagation of broad resonances (like $\rho$-meson) in the
medium. Decays of $\rho$-mesons are an important source of di-leptons radiated
by excited nuclear matter. As shown, a consistent description of the
invariant-mass spectrum of radiated di-leptons can be only achieved if one
accounts for the in-medium modification of the $\rho$-meson width (more
precisely, its spectral function).

We have argued that the Kadanoff--Baym equation within the 
first-order gradient approximation, slightly modified 
to make the set of Dyson's equations {\em exactly} consistent (rather than 
up to the second-order gradient terms), provide a proper frame for a quantum
four-phase-space kinetic description that applies also to systems of unstable 
particles. This quantum four-phase-space kinetic equation proves to be charge
and energy--momentum conserving and thermodynamically consistent, provided it 
is based on a $\Phi$-derivable approximation. The $\Phi$ functional also 
gives rise to a very natural representation of the collision term. 
Various self-consistent approximations are known since long time which do not
explicitely use the $\Phi$-derivable concept like self-consistent Born and
T-matrix approximations. The advantage the $\Phi$ functional method consists
in offering a regular way of constructing various self-consistent
approximations.

We have also addressed the question whether a closed
nonequilibrium system approaches the thermodynamic equilibrium during its
evolution. We obtained a definite expression for a local
(Markovian) entropy flow and were able to explicitly demonstrate the
$H$-theorem for some of the common choices of $\Phi$ approximations.  This
expression holds beyond the quasiparticle picture 
and thus generalizes the well-known Boltzmann kinetic
entropy.  Memory effects in the quantum four-phase-space kinetics were
discussed and a general strategy to deduce   
memory corrections to the entropy was outlined.\\ 

\noindent
{\bf Acknowledgment:}
We are grateful to G. Baym, G.E. Brown, P. Danielewicz, H. Feldmeier, B.
Friman, E.E. Kolomeitsev and P.C. Martin for fruitful
discussions.  Two of us (Y.B.I. and D.N.V.) highly appreciate the hospitality
and support rendered to us at Gesellschaft f\"ur Schwerionenforschung and
by the Niels Bohr Institute.  This work has been supported in part by BMBF 
(WTZ project RUS-656-96). Y.B.I. acknowledges partial support of 
Alexander-von-Humboldt Foundation. 

\end{document}